\documentclass[lettersize,journal]{IEEEtran}
\usepackage{amsmath,amsfonts}
\usepackage{algorithmic}
\usepackage{algorithm}
\usepackage{array}
\usepackage[caption=false,font=normalsize,labelfont=sf,textfont=sf]{subfig}
\usepackage{textcomp}
\usepackage{stfloats}
\usepackage{url}
\usepackage{verbatim}
\usepackage{graphicx}
\usepackage{cite}
\usepackage{tikz}
\begin{document}

\title{A Partial Compress-and-Forward Strategy for Relay-assisted Wireless Networks Based on Rateless Coding}

\author{Weihang Ding,~\IEEEmembership{Student Member,~IEEE,} and Mohammad Shikh-Bahaei,~\IEEEmembership{Senior Member,~IEEE}

\thanks{Copyright (c) 2015 IEEE. Personal use of this material is permitted. However, permission to use this material for any other purposes must be obtained from the IEEE by sending a request to pubs-permissions@ieee.org. }
\thanks{Weihang Ding and Mohammad Shikh-Bahaei are with the Department of Engineering, King’s College London, London WC2R 2LS, U.K. (e-mail: weihang.ding@kcl.ac.uk, m.sbahaei@kcl.ac.uk).}
}


\IEEEpubid{ }

\maketitle

\author{
\IEEEauthorblockN{Weihang Ding\IEEEauthorrefmark{1}, and Mohammad Shikh-Bahaei\IEEEauthorrefmark{1}
                  }
                  
\IEEEauthorblockA{\IEEEauthorrefmark{1}Centre for Telecommunications Research, Department of Engineering, King's College London, London WC2R 2LS, UK.}
}



\maketitle

\begin{abstract}
In this work, we propose a novel partial compress-and-forward (PCF) scheme for improving the maximum achievable transmission rate of a diamond relay network with two noisy relays. PCF combines conventional compress-and-forward (CF) and amplify-and-forward (AF) protocols, enabling one relay to operate alternately in the CF or the AF mode, while the other relay works purely in the CF mode. As the direct link from the source to the destination is unavailable, and there is no noiseless relay in the diamond network, messages received from both relays must act as side information for each other and must be decoded jointly. We propose a joint decoder to decode two Luby transform (LT) codes received from both relays corresponding to the same original message. Numerical results show that PCF can achieve significant performance improvements compared to decode-and-forward (DF) and pure CF protocols when at least the channels connected to one of the relays are of high quality.
\end{abstract}

\begin{IEEEkeywords}
Partial compress-and-forward, rateless coding, relays.
\end{IEEEkeywords}

\section{Introduction}
\IEEEPARstart{R}{elay}-assisted communication is a promising technique for wireless point-to-point communication. When the direct link from the source to the destination is not sufficient for supporting reliable transmissions, deploying relays can help improve coverage and achieve transmission diversity \cite{Nosratinia}. Based on the operations performed at the relays, relay-assisted networks can be divided into amplify-and-forward (AF), decode-and-forward (DF), and compress-and-forward (CF) protocols \cite{Laneman}, \cite{Cover}. However, in both AF and DF, the quality of the channel between the source and the relay acts as a bottleneck for the overall performance. In CF \cite{Cover}, the relay does not wait for the original message to be decoded successfully but compresses the received information as a new codeword and forwards the compressed codeword to the destination.
It has already been shown in \cite{Kang} that CF is optimal for diamond relay networks with one source, one destination, one noisy relay, and one noiseless relay. Lee and Chung also show that the optimal performance of such diamond relay networks can also be achieved with a combination of DF and CF protocols \cite{Lee}. In \cite{Savard}, the authors focus on a two-way diamond relay network. They assume that one of the relays implements a lattice-based CF scheme and compare the overall rate when the other relay performs AF, DF, or CF, respectively.

CF exploits the correlation between signals received from different paths and uses side information for decoding. Therefore, Wyner-Ziv (WZ) coding \cite{Wyner-Ziv} is widely used in CF strategies for information compression at the relays \cite{Hu}, \cite{Simoens}. The compressed codeword also needs to be channel-encoded before it is further forwarded to the destination. For simplicity, it is shown in \cite{Uppal1} that a channel coding scheme can be used for both information compression and error correction. Multi-level low-density parity check (LDPC) codes are used in \cite{Wan-l} for efficient implementation of CF. In \cite{Blasco-Serrano}, the authors propose a joint source-channel coding (JSCC) scheme for CF, which requires only low-complexity relays and effectively utilizes bandwidth. Either double-protograph LDPC codes \cite{S.Liu}, \cite{Q.Chen} or rateless codes \cite{Uppal} can be applied for JSCC.

\IEEEpubidadjcol

We propose a new scheme, Partial-CF (PCF), for utilizing a diamond relay network with two noisy relays. Conventional works typically assume the existence of a noiseless relay or direct link for CF \cite{Kang, Lee}. With two noisy relays, the codeword received from both relays cannot be assured to be error-free after quantization \cite{Nehra, Yirun1}. Therefore, CF is challenging to implement since no reliable side information is available at the destination. We propose a joint decoder to decode the compressed information from both relays, and we apply Luby Transform (LT) codes for JSCC in this correspondence. Operating channel-decoding and source-decoding separately leads to higher complexity and delay \cite{Bobarshad}. With JSCC, the information contents passing via the edges do not need to propagate among four decoders, which significantly speeds up the joint decoding at the receiver.
We realize that different compression rates lead to inefficiencies due to varying transmission times of the relays. Therefore, we propose the PCF scheme to more efficiently utilize the channels.
The major contributions of this work are listed as follows:
\begin{itemize}
\item We propose a novel PCF strategy for a diamond relay network with two noisy relays, which combines AF and CF to efficiently utilize the channels.
\item We optimize the timeline and compression rate at both relays for optimal performance in our PCF scheme.
\item We propose and implement a joint decoder for practical evaluation of our proposed PCF scheme by decoding the messages received from both relays where the information received through each relay is considered as side information for the other.
\end{itemize}

We compare the theoretical performance limits and the practical results obtained through simulations using LT codes for channel coding of our proposed PCF strategy with conventional protocols. The results indicate that applying PCF can provide a significant transmission rate gain when the channels are good. Even when some channels are bad, the performance can still be improved as long as the channels connected to one of the relays are good.

\section{System Model}

\begin{figure}[t!]
    \centering
    \includegraphics[width=0.55\columnwidth]{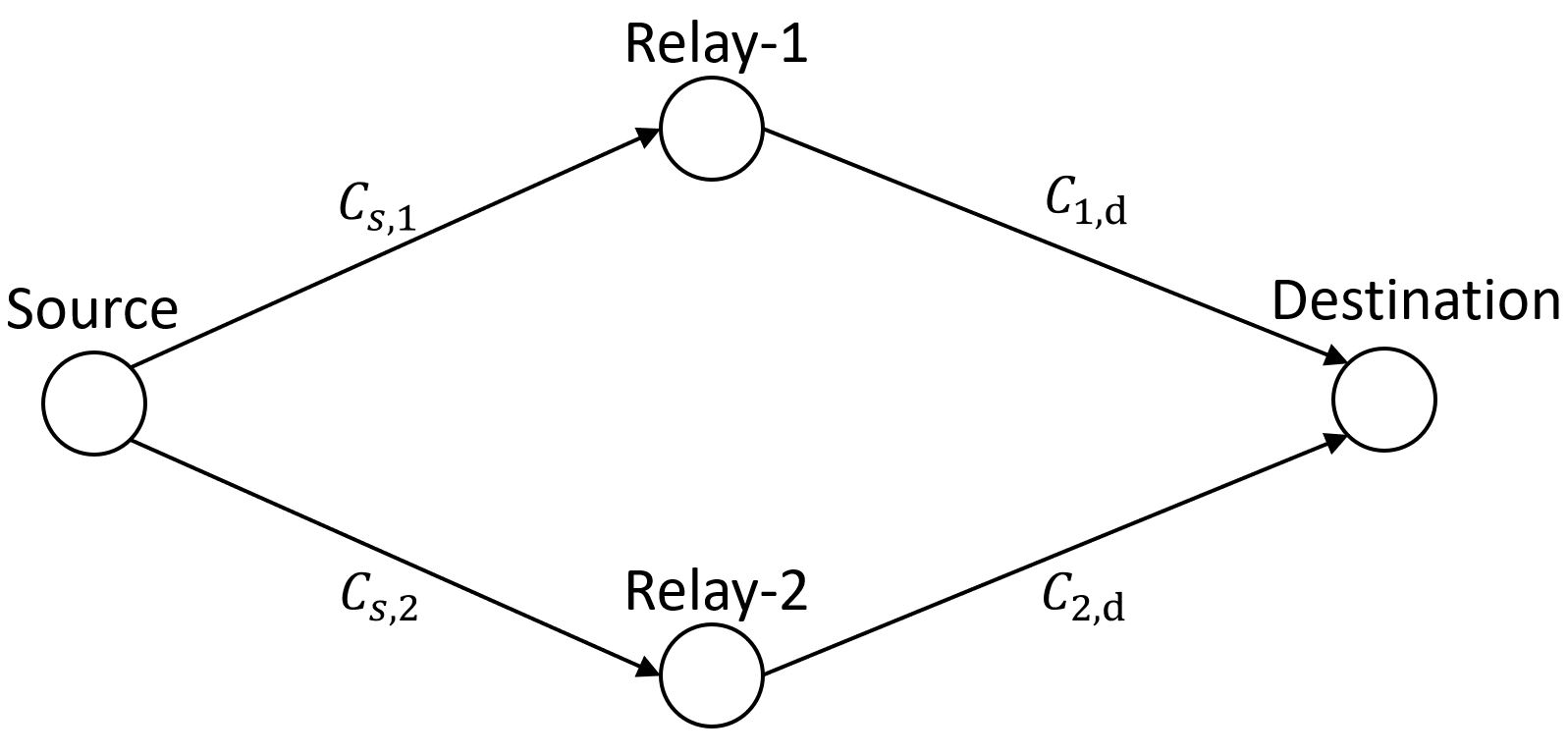}
    \caption{The wireless network model in this work consists of one source. one destination, and two noisy relays. The direct link from the source to the destination is unavailable.}
    \label{System_model}
\end{figure}

In reality, there might be multiple relays available in the network. However, it has been shown in \cite{Beres} that deploying multiple relays simultaneously is less efficient than deploying only the optimal relay. Therefore, we consider a diamond relay network with one source, two noisy relays, and one destination, as shown in Fig. \ref{System_model}. The relays are half-duplex, which means that they cannot receive and transmit messages simultaneously. It is also assumed that the direct link from the source to the destination does not exist. Let $C_{s,1}$, $C_{s,2}$, $C_{1,d}$, and $C_{2,d}$ represent the capacities of the channels between the source and Relay-1, between the source and Relay-2, between Relay-1 and the destination, and between Relay-2 and the destination, respectively. The channels from both relays to the destination are assumed to be orthogonal.

In this work, we assume that the source message is binary and transmitted using rateless codes. We define the original codeword to be transmitted through this diamond relay network as $\mathbf{Z}$, the compressed codeword at Relay-1 and Relay-2 as $\mathbf{X}$ and $\mathbf{Y}$, and the observations of $\mathbf{X}$ and $\mathbf{Y}$ at the destination as $\mathbf{U}$ and $\mathbf{V}$, respectively. As there is no message exchange between the relays, $\mathbf{X}$ and $\mathbf{Y}$ are conditionally independent given $\mathbf{Z}$. In addition, we denote $R_\mathbf{X}$ and $R_\mathbf{Y}$ as the average numbers of bits in $\mathbf{Z}$ that are compressed in each symbol of $\mathbf{X}$ and $\mathbf{Y}$, respectively. As quantization is deterministic at both relays, $R_\mathbf{X}$ and $R_\mathbf{Y}$ are bounded by the well-known Slepian-Wolf theory \cite{Slepian-Wolf}. We assume that the modulation and compression schemes used at both relays are the same, hence the length of codeword $\mathbf{X}$ ($\mathbf{Y}$) is inversely proportional to $R_\mathbf{X}$ ($R_\mathbf{Y}$). If the bandwidth allocated to both relays is the same, significant inefficiencies arise as one relay needs to wait for the other to complete its transmission.

\begin{figure}[t!]
    \centering
    \includegraphics[width=0.7\columnwidth]{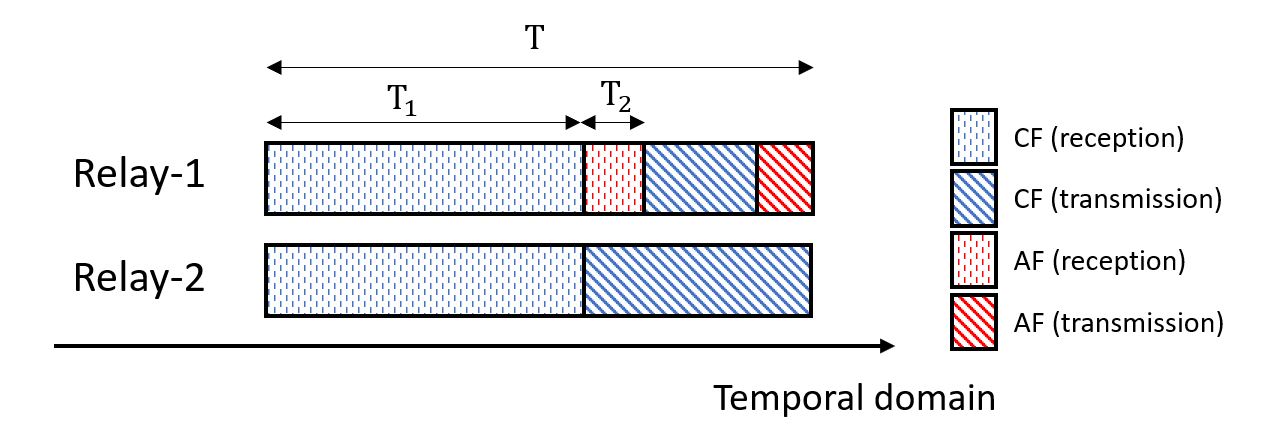}
    \caption{PCF in time domain. In this figure, we assume that Relay-2 works in the CF mode only, while Relay-1 works alternately on the CF and the AF mode.}
    \label{CF_temporal}
\end{figure}

To prevent relays from waiting during transmission, we propose a PCF strategy that combines the benefits of CF and AF protocols. We chose to use AF instead of DF in our design because the time allocated for this period is short and it is unlikely for the relay to successfully decode the packet in such a brief period of time. Here, we assume that $R_\mathbf{X}\geq R_\mathbf{Y}$, which means that $\mathbf{X}$ is shorter than $\mathbf{Y}$. Fig. \ref{CF_temporal} shows a sketch of this PCF protocol in the temporal domain. The source broadcast the signal for $T_1+T_2$ seconds in total. After the source broadcasts the original packet for $T_1$ seconds, Relay-2 stops listening and compresses the received signal into a longer signal before forwarding it for the remaining time. Meanwhile, Relay-1 continues to listen to the source for an additional $T_2$ seconds. The signal received during this period will be amplified and transmitted after transmitting the shorter compression of the signal received in the first $T_1$ seconds.

\begin{figure}[t!]
    \centering
    \includegraphics[width=0.75\columnwidth]{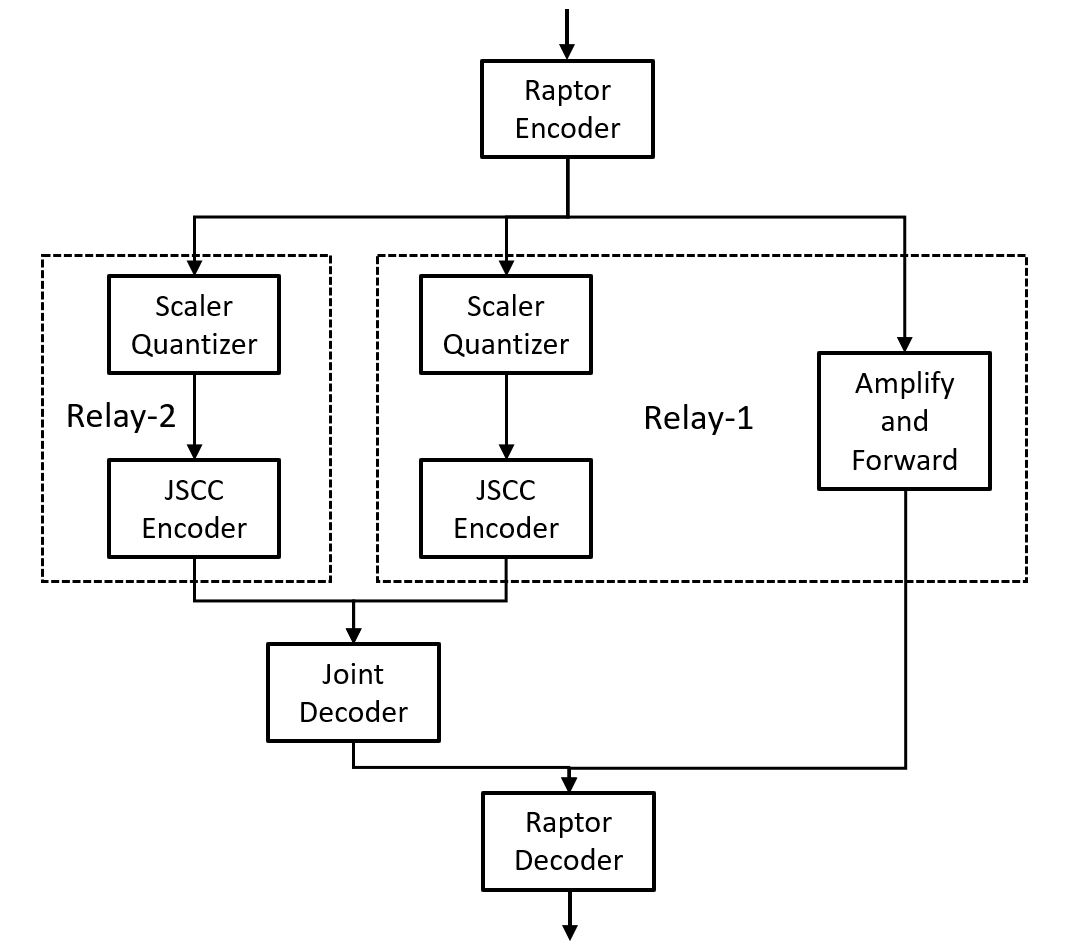}
    \caption{The block diagram of PCF strategy in this correspondence. We use Raptor codes at the source and LT codes as JSCC at both relays.}
    \label{BlockDiagram}
\end{figure}

In conventional CF protocols, the direct link provides important side information to the destination. However, as the direct link is unavailable in this scenario, the messages received from both relays should be used as each other's side information for decoding. The block diagram of PCF is shown in Fig. \ref{BlockDiagram}. For clarity, the original codeword is divided into two parts in the block diagram: one for CF and one for AF. However, in reality, the source does not need to split the codeword as the relays will autonomously switch to other modes after listening for a certain amount of time. After encoding and modulating, the resulting compressed codeword is forwarded to the destination. The compressed codewords received from both relays are jointly decoded, with each other serving as side information. Finally, the combination of both codewords is used for decoding and recovery of the original message.

\section{Capacity of PCF}

In this section, we compute the achievable transmission rate of our PCF scheme. Since we do not consider bandwidth allocation in this work, we assume that the same amount of bandwidth is allocated for transmission at all relays. After accumulating sufficient mutual information from the broadcast signal, one relay stops listening, compresses the received message, and forwards it during the remaining time of this transmission, while the other relay remains in listening mode for a period of time. Then, it compresses the message received in the first phase into a shorter codeword and forwards it, followed by the amplification of the messages received in the second phase.

The original message of length $k$ is divided into a codeword of length $k_1$ for CF and a codeword of length $k_2$ for AF. The observation of the first codeword of length $k_1$ is compressed into the codeword $\mathbf{X}$ of length $(k_1/R_\mathbf{X})$ at Relay-1 and the codeword $\mathbf{Y}$ of length $(k_1/R_\mathbf{Y})$ at Relay-2. The destination first tries to jointly decode $\mathbf{U}$ and $\mathbf{V}$, then combines the outputs of the joint decoder with the length-$k_2$ signal received from AF for final decoding.

Here, we define $\alpha_1=T_1/T$, and $\alpha_2=T_2/T$. $\mathbf{X}$ and $\mathbf{Y}$ can be recovered if the following conditions are satisfied:
\begin{equation}
    \alpha_1 R_X\leq (1-\alpha_1-2\alpha_2)C_{1,d},
    \label{pcf1}
\end{equation}
\begin{equation}
    \alpha_1 R_\mathbf{Y}\leq (1-\alpha_1)C_{2,d}.
    \label{pcf2}
\end{equation}

The original message with rate $R_{PCF}$ can be transmitted reliably through this network if
\begin{equation}
    R_{PCF}\leq \alpha_1I(\mathbf{Z};\mathbf{X},\mathbf{Y})+\alpha_2C_{1},
\end{equation}
where $C_1$ is the overall capacity of the source$\to$Relay-1$\to$destination link.

The upper bounds of $\alpha_1$ and $\alpha_2$ can be written as:
\begin{equation}
    \alpha_1\leq \frac{C_{2,d}}{R_{\mathbf{Y}}+C_{2,d}}
\end{equation}
\begin{equation}
    \alpha_2\leq \frac{C_{1,d}-\alpha_1(C_{1,d}+R_{\mathbf{X}})}{2C_{1,d}}
\end{equation}

\begin{figure}[t]
    \centering
    \resizebox{0.3\textwidth}{!}{%
    \begin{tikzpicture}
    \draw [thick,->](0,0) -- (5.5,0);
    \draw [thick,->](0,0) -- (0,5.5);
    \fill [gray!50](1,5.5) -- (1,3.5) -- (3.5,1) -- (5.5,1) -- (5.5,5.5);
    \draw [thin](0,1) -- (5.5,1);
    \draw [thin](0,3.5) -- (5.5,3.5);
    \draw [thin](1,0) -- (1,5.5);
    \draw [thin](3.5,0) -- (3.5,5.5);
    \draw [thin](4.5,0) -- (0,4.5);
    \draw [thin](1,5.5) -- (1,3.5) -- (3.5,1) -- (5.5,1);
    \filldraw[black] (1,0) circle (0.1pt) node[anchor=north] {$H(\mathbf{X}|\mathbf{Y})$};
    \filldraw[black] (3.5,0) circle (0.1pt) node[anchor=north] {$H(\mathbf{X})$};
    \filldraw[black] (4.7,0) circle (0.1pt) node[anchor=north] {$H(\mathbf{X},\mathbf{Y})$};
    \filldraw[black] (0,1) circle (0.1pt) node[anchor=east] {$H(\mathbf{Y}|\mathbf{X})$};
    \filldraw[black] (0,3.5) circle (0.1pt) node[anchor=east] {$H(\mathbf{Y})$};
    \filldraw[black] (0,4.5) circle (0.1pt) node[anchor=east] {$H(\mathbf{X},\mathbf{Y})$};
    \filldraw[black] (5.8,0) circle (0.1pt) node[anchor=north] {$R_\mathbf{X}$};
    \filldraw[black] (0,5.5) circle (0.1pt) node[anchor=east] {$R_\mathbf{Y}$};
    \filldraw[black] (0,0) circle (0.1pt) node[anchor=east] {$0$};
    \end{tikzpicture}
    }
    \caption{The joint admissible region of $R_\mathbf{X}$ and $R_\mathbf{Y}$ corresponds to our network defined by the Slepian-Wolf theory.}
    \label{Slepian-Wolf}
\end{figure}
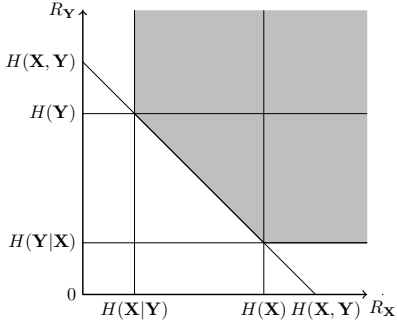

As $R_{PCF}$ is monotonically increasing with increasing $\alpha_2$, the optimization problem can be written as follows:

\begin{equation}
\begin{aligned}
\max_{R_\mathbf{X},R_\mathbf{Y},\alpha_1}&\alpha_1I(\mathbf{Z};\mathbf{X},\mathbf{Y})+\left(\frac{1}{2}-\frac{\alpha_1(C_{1,d}+R_{\mathbf{X}})}{2C_{1,d}}\right) C_1\\
\text{s.t.:}\:\:&\alpha_1(R_{\mathbf{Y}}+C_{2,d})\leq C_{2,d}\\
&R_\mathbf{Y}\leq R_\mathbf{X}\\
&(R_\mathbf{X},R_\mathbf{Y})\in \mathcal{R},
\end{aligned}
\label{opt}
\end{equation}
where $\mathcal{R}$ (shown in Fig. \ref{Slepian-Wolf}) is the joint admissible rate region of $R_\mathbf{X}$ and $R_\mathbf{Y}$, which is limited by the Slepian-Wolf bound \cite{Slepian-Wolf}. Problem \ref{opt} is a convex, but nonlinear optimization problem. The objective function is a non-increasing function of $R_\mathbf{X}$, regardless of the value of $R_\mathbf{Y}$. However, selecting the optimal value for $R_\mathbf{Y}$ requires knowledge of the channel state information (CSI) of the link between Relay-1 and the destination. For constant channels, obtaining this information is straightforward for Relay-2. However, if the channel is time-varying, the CSI needs to be delivered to Relay-2 from the destination through feedback before making the decision. The optimization problem can be efficiently solved using the Projected Gradient Descent (PGD) technique. Once the optimization problem is solved, the length and time duration of each period are determined. As it might be difficult to determine which relay should operate purely in CF mode and which relay should operate alternatively in CF or AF mode, we recommend solving the optimization problem twice and exchanging Relay-1 and Relay-2 in the second iteration to observe which policy yields better performance.

\section{Joint decoder}

\begin{figure}[t!]
    \centering
    \includegraphics[width=0.75\columnwidth]{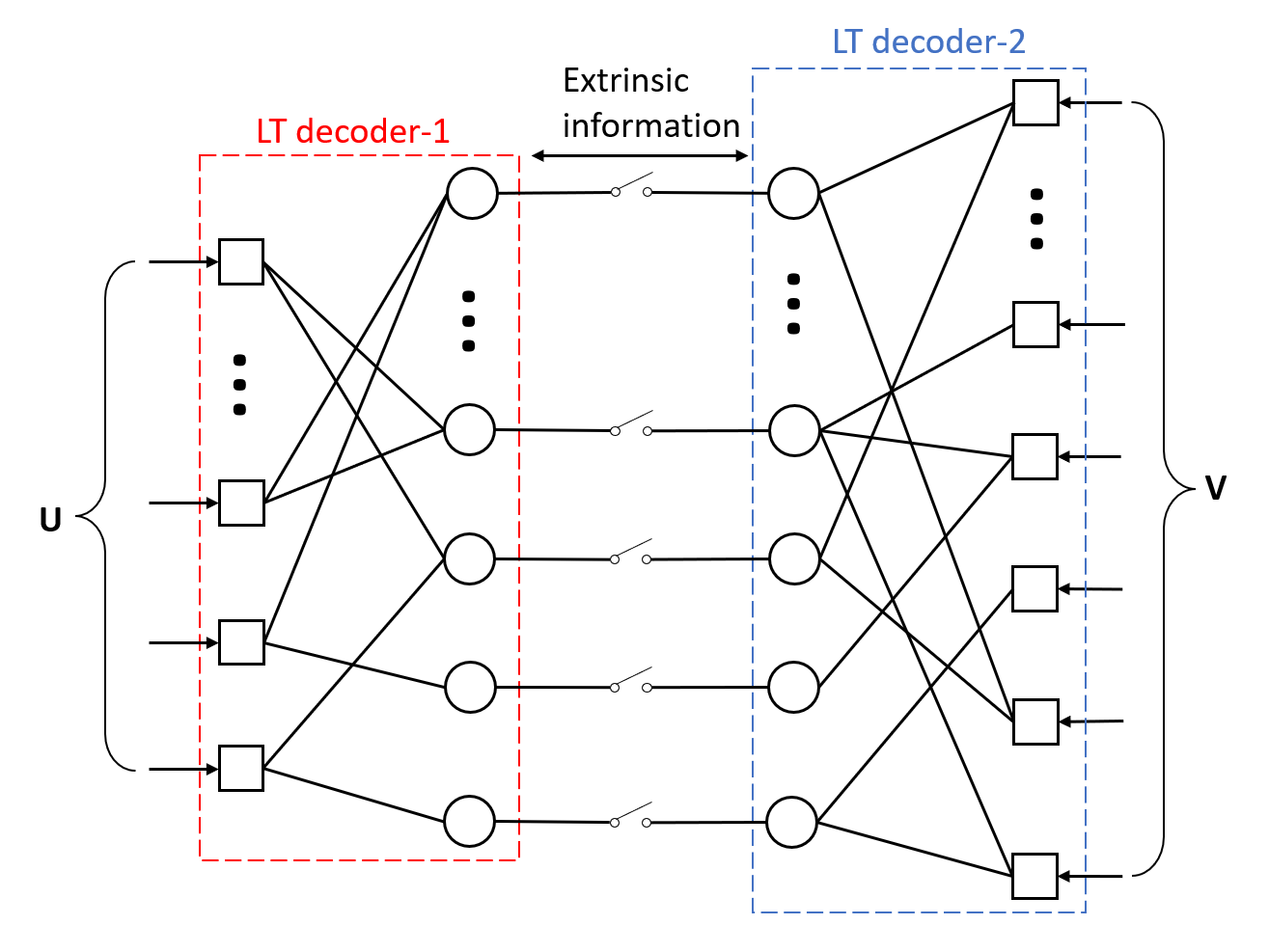}
    \caption{The joint decoder is represented by the Tanner graph. In the graph, circles represent variable nodes and squares represent parity-check nodes. The joint decoder is the concatenation of two belief-propagation decoders.}
    \label{Joint decoding}
\end{figure}

In our design, we use Raptor codes at the source to broadcast the message, but the signal received by the relay needs to be sampled and discretized for future processing. Even with a high-resolution scalar quantizer, some information is still lost, and using a high-resolution quantizer would make the compressed codeword much longer, which is not feasible in our system. Binary quantization has been shown to be superior in \cite{Uppal1} because the quantization thresholds do not need to be changed for different channel qualities. Therefore, we use a binary quantizer and a rateless encoder for joint source-channel coding at each relay.
While Raptor codes are also used for compression in \cite{Uppal}, we found that they lead to some extra problems in our work. The length of the received rateless code varies in different transmissions due to time-varying channels, which makes it difficult to prepare a sufficient number of base graphs for different block sizes if we want to use LDPC codes for precoding. Additionally, Raptor codes are more time-consuming and complicated to decode than LT codes, making them unsuitable for our joint decoding algorithm. Therefore, we only consider LT codes as the joint source-channel coding method in our work.

The joint decoding algorithm is described in Fig. \ref{Joint decoding}.
Both LT decoder-1 and LT decoder-2 are belief-propagation (BP) decoders. In each decoding iteration, the decoding is based on the channel inputs and the extrinsic information from the other decoder, which serves as side information. The output extrinsic information is expressed in terms of log-likelihood radios (LLRs). Let us denote the $i$-th output LLR value of LT decoder-1 after the $l$-th decoding iteration by $r_{\mathbf{X}\to \mathbf{Y},i}^{(l)}$, and the $i$-th bit of $\mathbf{Y}$ by $y_i$. The probability that $y_i=1$ given $r_{\mathbf{X}\to \mathbf{Y},i}^{(l)}$ can be calculated by:
\begin{equation}
\begin{aligned}
    &P(y_i=1|r_{\mathbf{X}\to \mathbf{Y},i}^{(l)})=\\
    &(P_{e,1}P_{e,2}+(1-P_{e,1})(1-P_{e,2}))\frac{r_{\mathbf{X}\to \mathbf{Y},i}^{(l)}}{1+r_{\mathbf{X}\to \mathbf{Y},i}^{(l)}}\\
    &+((1-P_{e,1})P_{e,2}+(1-P_{e,2})P_{e,1})\frac{1}{1+r_{\mathbf{X}\to \mathbf{Y},i}^{(l)}},
\end{aligned}
\label{P1}
\end{equation}
where $x_i$ is the $i$-th bit of $\mathbf{X}$, $P_{e,1}$ ($P_{e,2}$) is the bit-error-rate of the channel between the sensor and Relay-1 (Relay-2). Similarly, the probability of $y_i=0$ given $r_{\mathbf{X}\to \mathbf{Y},i}^{(l)}$ can be calculated as:
\begin{equation}
\begin{aligned}
    &P(y_i=0|r_{\mathbf{X}\to \mathbf{Y},i}^{(l)})=\\
    &((1-P_{e,1})P_{e,2}+(1-P_{e,2})P_{e,1})\frac{r_{\mathbf{X}\to \mathbf{Y},i}^{(l)}}{1+r_{\mathbf{X}\to \mathbf{Y},i}^{(l)}}\\
    &+(P_{e,1}P_{e,2}+(1-P_{e,1})(1-P_{e,2}))\frac{1}{1+r_{\mathbf{X}\to \mathbf{Y},i}^{(l)}}.
\end{aligned}
\label{P2}
\end{equation}

Using $P(y_i=1|r_{\mathbf{X}\to \mathbf{Y},i}^{(l)})$ and $P(y_i=0|r_{\mathbf{X}\to \mathbf{Y},i}^{(l)})$, the $i$-th input LLR value of LT decoder-1 after the $l$-th iteration $\iota_{\mathbf{X}\to \mathbf{Y},i}^{(l)}$, can be computed as:

\begin{equation}
    \iota_{\mathbf{X}\to \mathbf{Y},i}^{(l)}=\log\left(\frac{P(y_i=0|r_{\mathbf{X}\to \mathbf{Y},i}^{(l)})}{P(y_i=1|r_{\mathbf{X}\to \mathbf{Y},i}^{(l)})}\right),
\end{equation}
and $\iota_{\mathbf{Y}\to \mathbf{X},i}^{(l)}$ can be calculated similarly. These LLRs are then used by the other BP decoder as side information. Ideally, this decoding algorithm will not terminate until both decoders have successfully recovered their message or a time-out occurs. However, to save time, we set an upper bound for the operating iterations in practice.

\section{Numerical Results}

In this section, we compare the performance of our proposed PCF scheme with conventional AF and DF schemes, as well as the capacity achieved by the best relay in a diamond relay network \cite{Jain}. In the simulations, we assumed that BPSK is applied at the source and both relays. The source used Raptor codes to encode the original information of 4000 bits, with the Raptor code consisting of a rate-0.95 (3,60) regular LDPC code and an LT code. The degree distribution of the LT code was determined using the method proposed in \cite{Ravanshid}. The joint decoder was allowed a maximum of 40 decoding iterations. In the DF protocol, the relays and destination attempted to recover the message every time they received 100 additional bits. In PCF and pure CF, the time at which the relay should stop listening and the length of the compressed codeword were predetermined based on the channel qualities with the achievable rate being computed when at least 99\% reliability was ensured, and the LT codes received from both links were jointly decoded at the destination. The values of $R_\mathbf{X}$, $R_\mathbf{Y}$, and the duration of each phase were determined to ensure reliable transmissions. In the single-relay scenario, we assign twice the bandwidth to the relay compared to the double-relay scenario to ensure fair comparisons.

\begin{figure}[t!]
    \centering
    \includegraphics[width=0.7\columnwidth]{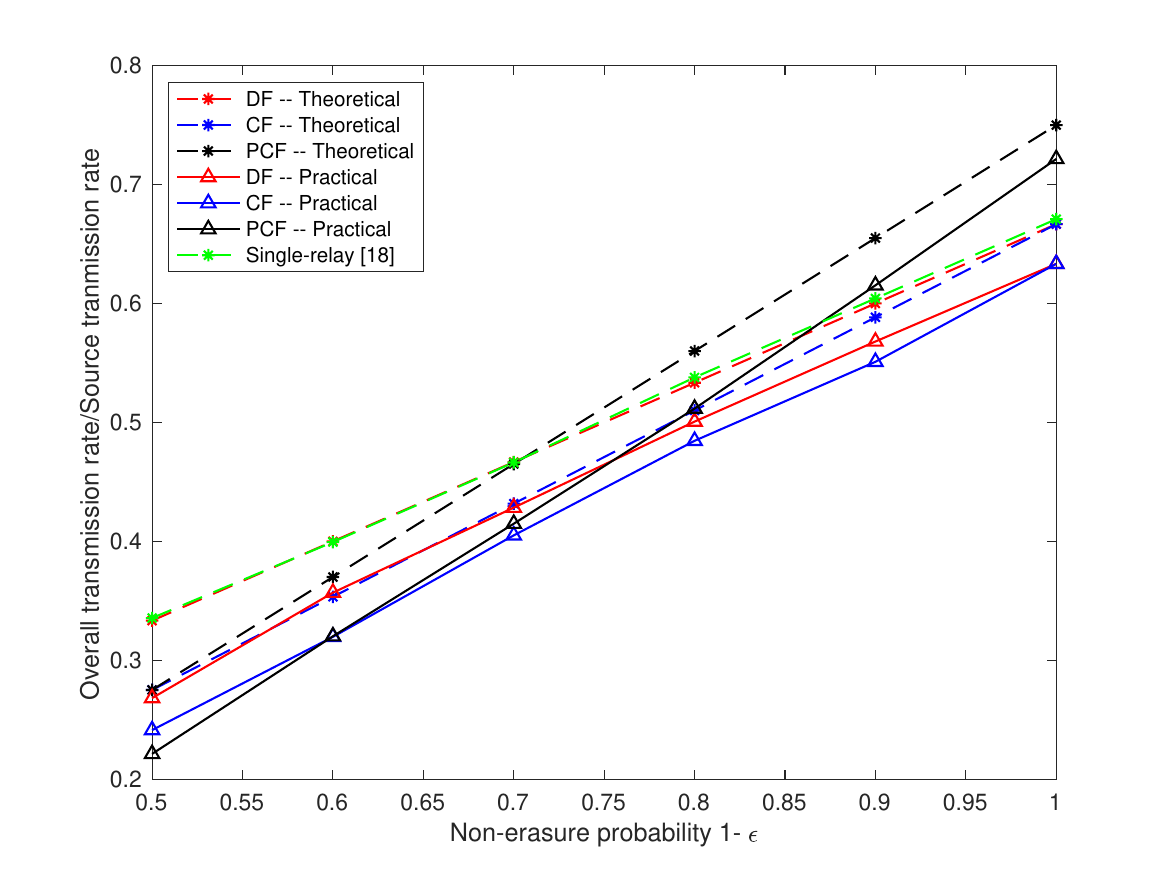}
    \caption{The transmission rate vs. non-erasure probability $(1-\epsilon)$. The transmission rates are compared according to their ratios to the original transmission rate at the source. It is assumed that the channels in the network are BECs with identical erasure probabilities.}
    \label{BEC_capacity}
\end{figure}

At first, we modeled all channels in the network as binary erasure channels (BECs). Since there are four independent channels in the network, we assumed for simplicity that the channel conditions of all four channels remained the same at all times. While this assumption is unnecessary in reality, it makes the results more intuitive in the figures. Fig. \ref{BEC_capacity} shows the achievable rates and the actual simulation results as functions of the non-erasure probability ($1-\epsilon$). The achievable rates are plotted as ratios to the original transmission rate at the source. Our proposed PCF scheme outperforms all other schemes in terms of transmission rate when the channel qualities are better. The simulation results show similar performance, although some degradation is unavoidable due to finite codeword length and the rate-0.95 predcoding.

\begin{figure}[t!]
    \centering
    \includegraphics[width=0.7\columnwidth]{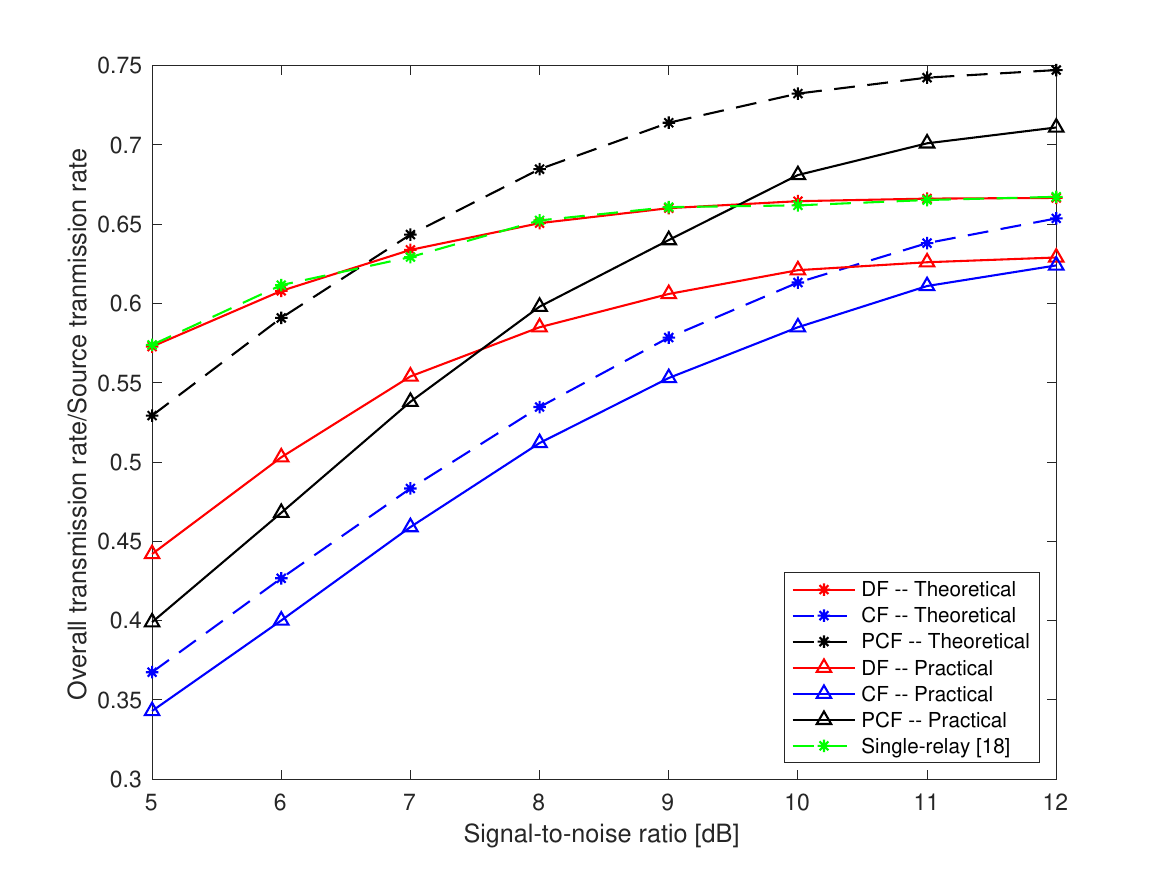}
    \caption{The overall transmission rate vs. SNR. Because of binary quantization at the relays, the channels between the source and the relays are modeled as BSC channels.}
    \label{BIAWGN_capacity}
\end{figure}

We then assume that all the channels are binary additive white Gaussian noise (BI-AWGN) channels. Despite being transmitted through BI-AWGN channels, the output of the binary quantizer is not actually the received signal, but instead, ``1"s and ``0"s. We use a single-threshold hard-decision quantizer, which maps positive LLRs to ``0" and negative LLRs to ``1". Therefore, the channels between the source and the relays are modeled as binary symmetric channels (BSCs) with crossover probabilities determined by the bit-error-rates (BERs) of the BI-AWGN channels. The results in Fig. \ref{BIAWGN_capacity} show that in lower SNR regimes, DF performs better. However, when the SNR of all the channels is above 6.7 dB, applying PCF can result in higher achievable rates compared with other schemes. In practice, PCF outperforms DF at around 7.5 dB SNR. PCF is not as effective as DF when the channels are of low quality because noise enhancement is still significant in PCF. However, applying PCF can significantly improve the upper bound of performance compared with DF.

\begin{figure}[t!]
    \centering
    \includegraphics[width=0.7\columnwidth]{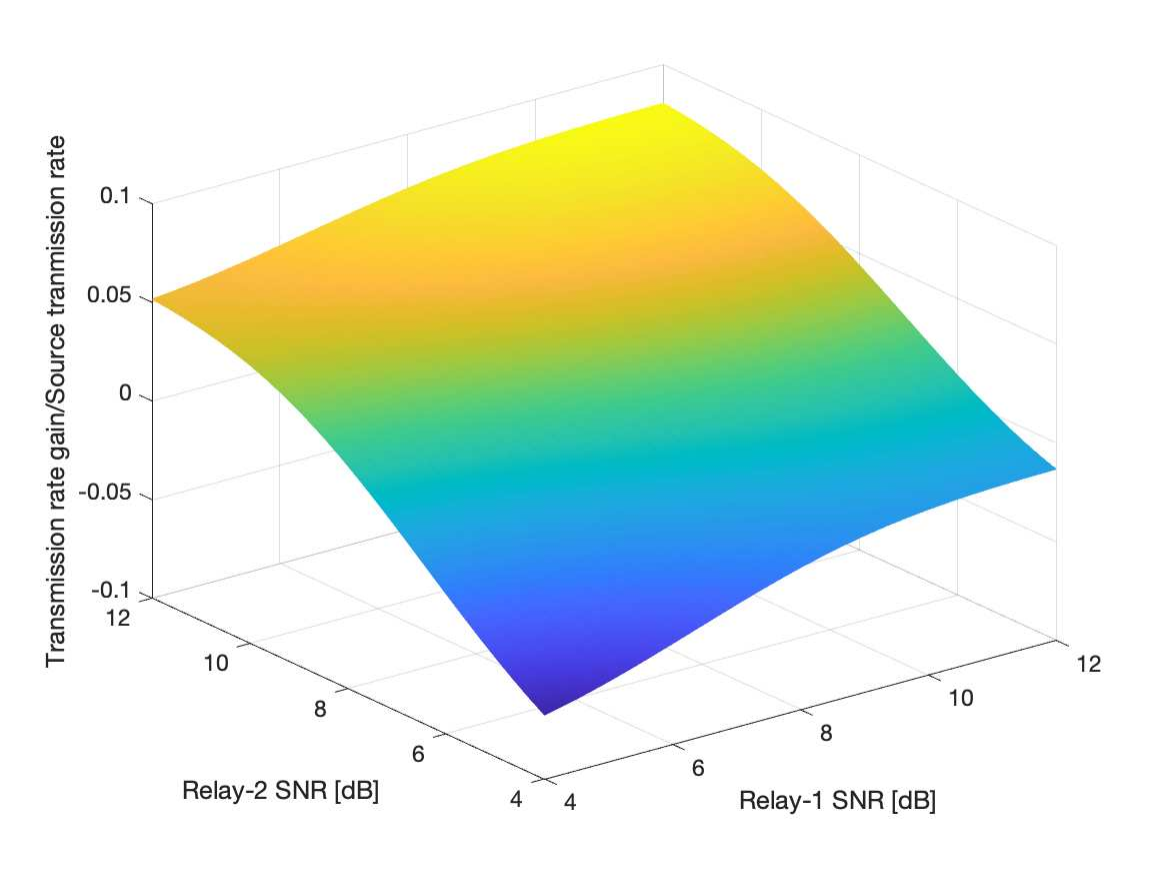}
    \caption{The transmission rate gain of PCF compared with DF, for different channel conditions. The lighter the colour, the more we can gain from applying PCF.
    }
    \label{PCFvsDF}
\end{figure}

From the above results, we can conclude that PCF is better only when all channels in the network are of high quality. However, we also find that PCF performs well when only the channels connected to one of the relays are good. The performance gain of partial-CF compared with DF is shown in Fig. \ref{PCFvsDF}. Again, here we assume that the channel qualities of both channels connected to each relay are the same. The SNR of the pure CF relay, i.e., Relay-2, is more significant than that of Relay-1. When the SNRs of the channels connected to Relay-2 are high, applying PCF can bring us performance gain compared with other protocols regardless of the qualities of the channels connected to Relay-1. Therefore, the relay with better channels should always be selected to work purely in the CF mode.

\section{Conclusion}
In this work, we show how to deploy the CF protocol in a diamond relay network with two noisy relays. We propose a PCF strategy that can improve the overall transmission rate when the channels are of high quality. This conclusion is not just achieved theoretically; we also conduct simulations to show that PCF works in practice. We design and deploy a joint LT code decoder at the destination to decode the signal received from both relays, resulting in similar performance gains to the theoretical results. The practical performance in can be further improved by increasing the length of the codewords or using more powerful JSCC schemes.

\bibliographystyle{IEEEtran}
\bibliography{IEEEabrv,Ref}

\vfill

\end{document}